\documentclass[openacc]{rstransa}

\newcommand*{\doi}[1]{\href{http://dx.doi.org/#1}{doi: #1}}

\titlehead{Research}

\begin{document}

\title{Generalized Geometric Brownian motion and the Infinite Ergodicity
concept}

\author{
S. Giordano$^{1}$ and R. Blossey$^{2}$}


\address{$^{1}$ University of Lille, CNRS, Centrale Lille, Univ. Polytechnique Hauts-de-France, UMR 8520 - IEMN - Institut d'{\'E}lectronique, de Micro{\'e}lectronique et de Nanotechnologie, F-59000 Lille, France\\
$^{2}$ University of Lille, Unit{\'e} de Glycobiologie Structurale et Fonctionnelle (UGSF), CNRS UMR8576, F-59000 Lille, France}

\subject{Stochastic processes, infinite ergodicity}


\corres{Ralf Blossey\\
\email{ralf.blossey@univ-lille.fr}}

\begin{abstract}

We investigate stochastic processes that generalize geometric Brownian motion, focusing on cases where the standard invariant measure—i.e., the solution of the stationary Fokker–Planck equation—does not necessarily exist. We demonstrate that the existence of such a measure depends sensitively on the structure of the drift and diffusion terms, as well as on the chosen discretization scheme of the underlying stochastic dynamics. To ground our discussion, we draw motivation from phenomenological models in statistical theories of turbulence, where geometric Brownian motion serves as a classical example. To address situations where the standard invariant measure fails to exist, we heuristically explore the concept of infinite ergodicity, a notion recently introduced in the context of statistical physics for drift–diffusion stochastic processes.

\end{abstract}


\begin{fmtext}
\section{Introduction}


Geometric Brownian Motion (GBM) is a stochastic process
whose solution of its Fokker-Planck equation is given by the 
log-normal distribution. GBM is {\it the} fundamental stochastic process of mathemetical finance, as it underlies the famous Black-Scholes formula of stock option pricing
in which it relates to the logarithmic return or profit of a stock prize.
In the finance context, and more specifically within the statistical
physics (``phynance") literature, it has found numerous 
modifications and generalizations, see, e.g., Refs. \cite{stojkoski20,peng24}.
\end{fmtext}
\maketitle
In the context of applying GBM to finance, discussions have also emerged concerning the notion of ergodicity, a key concept in equilibrium statistical physics \cite{peters13}. 
In this paper we wish to place our discussion of GBM and some  nonlinear generalizations in the context of a seemingly less discussed topic with respect to geometric Brownian motions: turbulence. The notion of ergodicity
gains a more physical relevance in this context. 

The statistical nature of the turbulence problem has been noticed long ago, and a large body of literature has been built on this basic concept. 
Of course, we can only touch upon this topic, as a full discussion lies outside the scope of this paper. 
The following introductory remarks 
are inspired by the recent monograph by Birnir \cite{birnir13}, in which the author develops a mathematical approach to turbulence around the Kolmogorov-Obukhov phenomenological scaling theory. 
In turbulence, velocity increments or energy transfer across scales exhibit strong randomness that cannot be captured by purely deterministic models.
Therefore, in 1941, the two Russian scientists independently postulated a statistical theory of turbulence based on phenomenological arguments, postulating 
power-law behaviour for the structure function of velocity differences
\cite{kolmogorov41,obukhov41}.
Following criticisms, the original theory was corrected in 1962 to take into account large flow structures and intermittency effects, leading to log-normal corrections \cite{kolmogorov62,obukhov62}. 
Thus, here they arise 
the log-normal effects - and the geometric Brownian motion theory comes into play. In his monograph, Birnir derives the Kolmogorov-Obukhov theory from the stochastic Navier-Stokes equation with a judicious choice of the noise \cite{birnir13}. Interested readers are recommended to follow this journey 
there. There is, however, a mathematically simpler way to arrive at the log-normal behaviour in turbulence, described by geometric Brownian motions. This path can be followed via experiments.

In a series of papers written over several decades, J. Peinke and collaborators have developed an approach to describe the turbulent energy cascade by Markov processes, with the description based on a Langevin equation for the trajectories of velocity increment
across the spatial scales of the turbulent fluid \cite{friedrich97,naert97,fuchs20}. The theory is developed on the basis of experiments, e.g. the hotwire velocity measurements $v(t)$,
where $v$ is the velocity component in the direction of mean flow \cite{fuchs22}. The experiments allow to determine the longitudinal velocity increments at scale $r$
\begin{equation}
    u_r = v(t+\tau) - v(t) 
\end{equation}
based on the scale hypothesis $r = -\tau \langle v \rangle $.
The probability distribution function (PDF) is given by 
$P(u,r) = \mathbb{E}[\delta(u_r - u)]$, as a function of scale $r$. Performing a change to the logarithmic variable $s = -\ln(r/L)$ 
between an initial value $s_i = 0$ at $r=L$, and a final positive value $s_f = -\ln(\lambda/L)$ at the Taylor scale $\lambda$, one finds cascade trajectories $u_s$. The stochastic process $u_s$ in cascade scale has been found to
be Markovian down to a scale $\approx 0.9 \lambda$ and can
be modeled as a diffusion process, written in the notation of Ref. \cite{fuchs22}
\begin{equation} \label{peinke}
    du_s = D^{(1)}(u_s,s) ds + \left[2 D^{(2)}(u_s,s) \right]^{1/2} dW_s,
\end{equation}
where $W_s$ is a Wiener process. The equation, despite being written in the typical It\^o format, is considered in the
Stratonovich interpretation, i.e. with a discretization parameter $\alpha = 1/2$. Trajectories can be generated by numerical integration with the choice of drift and diffusion coefficients
\begin{equation} \label{peinke2}
    D^{(1)}(u,s) = -(\alpha_0 + \gamma_0) u\,,\,\,\,
    D^{(2)}(u,s) = \beta_0 + \gamma_0 u^2\, ,
\end{equation}
with an initial condition as a centered Gaussian distribution
fitted to experiment (note that we have added a subscript "0" to these parameters in order to avoid confusion with parameters introduced by us).
This approach allows to reproduce the improved scaling theory by Kolmogorov and Obukhov (K62) with the parameters
\begin{equation} \label{peinke3}
    \alpha_0 = (3 + \mu)/9\,,\,\, \beta_0 = 0\,,\,\, \gamma_0 =
    \mu/18\, 
\end{equation}
with the intermittency parameter $\mu = 0.234$ introduced by Kolmogorov; other empirical choices for the parameters are also discussed \cite{fuchs22}.
In the case $\beta_0 = 0$, the Langevin equation can be solved analytically, which can be shown in a fairly straightforward computation.

We now take the presence of the log-normal distribution, generated by the
geometric Brownian motion, i.e. a process with multiplicative noise as the starting point for a more general discussion of its properties. In fact,
the standard GBM does not generate a stationary PDF for large scales,
which is the result obtained for the assumed Markov process across spatial
scales. The existence of a stationary PDF is equivalent to the existence of an {\it invariant measure} or {\it invariant density}, from which equilibrium statistical averages can be obtained.

It should be noted that in constructing a phenomenological theory to be fitted to experimental data, one is invariably required to address measurement noise, see Refs. \cite{kleinhans07,kleinhans12}.
This may not only affect the parametrization of the model (see, e.g., Ref.\cite{fuchs22}),
but also the overall form of the equation maybe subject to scrutiny. Indeed, the traditional GBM cannot faithfully reproduce altogether the observed fat tails, the multi-fractal statistics, and the cascade mechanisms of turbulent velocity increments. Other models for turbulent phenomena notably make use of nonlinear diffusion terms, that are again also frequently used for finance problems \cite{cox85,heston93,bensoussan16}.
In the Langevin equation, while the drift term represents systematic tendencies, such as average energy transfer between scales, the diffusion term encodes random variability, reflecting the inherently chaotic nature of turbulent flow.
We will in the following therefore step over to a conceptual discussion, and consider Langevin equations for generalized geometric Brownian motions, in which both drift and diffusion coefficients deviate from the linear behaviour discussed so far, focussing exclusively on algebraic nonlinearities (e.g., the square root process).

Issues of (non-)ergodicity have recently e.g. been studied for fractional Brownian motion (FGBM) \cite{metz1,metz2,metz3}
and under stochastic resetting \cite{barkai23}.
The choice of stochastic calculus can influence predictions, echoing ambiguities in how noise interacts with nonlinear turbulent dynamics.
We will see that, in conjunction with different discretization rules of the stochastic process as given by the
parameter $ 0 \leq \alpha \leq 1 $, the existence of invariant measures can be answered in a novel way by invoking the concept of {\it infinite ergodicity} \cite{barkai23,leibovich19,aghion20,giordano23}. 
This approach finds physical applications to subrecoil laser cooling \cite{akimoto22,akimoto23}, to time-delayed feedback systems \cite{albers25}, and to intermittent chaotic dynamical systems \cite{okubo21,yan24}. The concept of infinite ergodicity generalizes the relation between time and space averages in equilibrium statistical mechanics whose identity is encoded in the notion of ergodicity, since the time average is not equivalent to the space average in infinite ergodic systems. The asymptotic behaviour of the average values of an arbitrary variable can, however, be determined as the ensemble average with appropriate invariant densities. In the context of turbulence in which we present our discussion, the notion of intermittency -- highly relevant in turbulent behaviour -- has been addressed in the infinite ergodic context particularly in discrete maps following Pomeau and Manneville \cite{pomeau80}. For a recent discussion of this large field, see \cite{nakagawa14,okubo21,yan24}.

\section{Generalized Geometric Brownian motions}

In a first step, we now generalize the stochastic process of geometric Brownian motion to processes with
a more general discretization rule, invoking the parameter $\alpha$
for which $0 \leq \alpha \leq 1$, but stay with the GBM first.
Further, we will render the notation a
bit lighter and replace eq. (\ref{peinke}) by the Langevin equation in a more classical notation as
\begin{equation} \label{gbm}
    \frac{du}{ds} = H(s)u(s) + G(s)u(s)\xi(s) 
\end{equation}
where we define the Wiener process by the Gaussian
random variable $\xi(s)$ with mean $\mathbb{E}{[\xi(s)]} = 0$
and correlation $\mathbb{E}[\xi(s)\xi(s')] = 2\delta(s-s')$. 
The Langevin eq. (\ref{gbm}) corresponds to the Fokker-Planck equation for the PDF $P(u,s)$ given by
\begin{equation}
    \frac{\partial}{\partial s}P(u,s) = -H(s)\frac{\partial}{\partial u}[uP(u,s)]
     + G^2(s)\frac{\partial}{\partial u}
     \left[u^{2\alpha}\frac{\partial}{\partial u} \left[u^{2(1-\alpha)}P(u,s)\right]\right]
\end{equation}
in which the discretization parameter $0 \leq \alpha \leq 1$
explicitly appears. Its main values are: $\alpha = 0$ (It\^o),
$\alpha = 1/2$ (Fisk-Stratonovich) and $\alpha = 1$ (H\"anggi-Klimontovich). 
Note that here we have generalized the prefactors of the linear terms to scale-dependent functions \cite{giordano23}. Further, while keeping the variable notation $s$ as introduced for the turbulent scale, it should
be clear that, depending on context, it can also serve as the time variable in the stochastic process.

The solution of this equation can be obtained from the following elegant argument. We assume that the driftless case $H\equiv0$ 
(with constant $G$) is solved by 
\begin{equation} \label{log-normal}
    f(u) = \frac{1}{u\sqrt{2\pi\sigma^2}}
    \exp{-\frac{(\log u - \overline{m})^2}{2\sigma^2}}, 
\end{equation}
with parameters $\overline{m}$ and  $\sigma^2$ 
(note that these parameters do not correspond to the mean value and variance).
For this distribution, the following relations hold
for the expectation values of $u$ and $u^2$:
\begin{equation}
    \mathbb{E}[u] = e^{\overline{m} + \frac{\sigma^2}{2}}\,,\,\,\,
    \mathbb{E}[u^2] = e^{2(\overline{m} + \sigma^2)} ,
\end{equation}
with the variance given by $\sigma_u^2 = \mathbb{E}[u^2] - \mathbb{E}[u]^2$. From the above relations, we can likewise
compute the parameters $\overline{m}$ and  $\sigma^2$ from the expectation values to find:
\begin{equation}
    \overline{m} = \log\left\{ \frac{\mathbb{E}[u]^2}{\sqrt{\mathbb{E}[u^2]}}\right\}\,,
    \,\,\,\, \sigma^2 = \log\left\{ \frac{\mathbb{E}[u^2]}{\sqrt{\mathbb{E}[u]^2}}\right\}\, .
\end{equation}
Assuming that log-normality also holds for arbitrary $s$-dependent
drift and diffusion terms, by multiplication of the Fokker-Planck equation with $u$ and $u^2$ and subsequent integration one obtains 
the two first-order differential equations for the expectation values
\begin{equation}
    \frac{d\mathbb{E}[u]}{ds} = [H(s) + 2\alpha G^2(s)] \mathbb{E}[u] ,
\end{equation}
and
\begin{equation}
    \frac{d\mathbb{E}[u^2]}{ds} = 2[H(s) + (2\alpha + 1) G^2(s)] \mathbb{E}[u] ,
\end{equation}
which can be solved by straightforward integration, e.g. on the 
interval $s = (0,\infty)$, and from the solutions we then obtain explicit expressions for the mean and the variance. We ultimately
find the expression for the log-normal distribution with the initial condition $ P[u,0] = \delta[u - \overline{u}_0]$ as in 
eq. (\ref{log-normal}) with
\begin{equation}
    \overline{m} = \log \overline{u}_0 +
    \int_0^{s}dt[H(t) + (2\alpha - 1)G^2(t)] 
\end{equation}
and
\begin{equation}
    \sigma^2 = 2\int_0^s dt G^2(t)\,, 
\end{equation}
whereby $\overline{u}_0 = \mathbb{E}[u](0) $ and $\sigma_u^2 (0)=0$ have been set. The validity of this result can be checked explicitly by calculation, for details, see Ref. \cite{giordano23}.
Our result generalizes the log-normal solutions to discretization values
$0 \leq \alpha \leq 1$ for arbitrary scale-dependent prefactors. 

In the limit $s \rightarrow \infty $, the existence of the distribution depends on the behaviour of the prefactors $H$ and $G$, so that the integrals
in the expression lead to meaningful finite limits. The following discussion is devoted to this issue, and we will also look at slightly generalized
geometric Brownian motions (GGBMs) given by the Langevin equation
\begin{equation}
    \frac{du}{ds} = h(u) + G_0u\xi(s)\,,
    \label{langGGBM}
\end{equation}
where we, for the moment, leave the drift term $h(u)$ unspecified.
We will be searching for the asymptotic solution of the corresponding
Fokker-Planck equation for large scales $s \rightarrow \infty$, which formally fulfills the first-order differential equation
\begin{equation}
    -h P_{\infty} + G_0^2u^{2\alpha}\frac{d}{du}[u^{2(1-\alpha)}P_{\infty}] = 0\, .
\end{equation}
Defining $\Pi(u) \equiv u^{2(1-\alpha)} P_{\infty}$ we can rewrite this expression as 
\begin{equation}
    \frac{d\Pi(u)}{du} = \frac{h(u)}{G_0^2u^2}\Pi(u)\, ,
\end{equation}
which can be readily integrated and leads to the expression 
\begin{equation} \label{P-ln}
    P_{\infty}(u) = \frac{K}{u^{2(1-\alpha)}}\exp\left(\int du \frac{h(u)}{G_0^2u^2}\right)\, ,
\end{equation}
where $K$ is a normalization constant. As we have pointed out in Ref.
\cite{giordano23}, this result bears a resemblance to the Pope-Ching result already invoked for turbulent flow data \cite{pope93}.

As discussed before, in the case of $h(u) = H_0 u$, the normalization
integral $\int_0^{\infty} du P_{\infty}(u)$ cannot converge to a finite value. Expression (\ref{P-ln}) then in fact yields the algebraic law
\begin{equation}
    P_{\infty}(u) = K u^{-2(1-\alpha)+ H_0/G_0^2},
\end{equation}
through which the integral expression either diverges at the lower or
upper limit, which follows from chopping up the integral limit into two
intervals $(0,1)$ and $(1,\infty)$. 
This is precisely the case of Eqs. (\ref{peinke}), (\ref{peinke2}) and (\ref{peinke3}), where we must consider $H_0=-1/3-\mu/6$, and $G_0=\sqrt{\mu/9}$. Therefore, we get $ P_{\infty}(u) = K u^{-2(1-\alpha)-3/2-3/\mu}$. If we implement the Stratonovich interpretation, we have the non-normalizable density $ P_{\infty}(u) = K u^{-5/2-3/\mu}$.

As a generalization of the geometric Brownian process we now consider a nonlinear drift of the form
\begin{equation}
    h(u) = -H_0u^n  
\end{equation}
in which case we have
\begin{equation} \label{P-ln2}
    P_{\infty}(u) = \frac{K}{u^{2(1-\alpha)}}\exp\left(-\frac{H_0}{G_0^2}\frac{u^{n-1}}{n-1}\right)\,, 
\end{equation}
for $n \neq 1$. In order to have a normalizable asymptotic density, the
inverse of the normalization constant must be given by the finite 
expression
\begin{equation}
\label{partinte}
    \frac{1}{K} = \int_0^{\infty} du \frac{1}{u^{2(1-\alpha)}}
    \exp\left(-\frac{H_0}{G_0^2}\frac{u^{n-1}}{n-1}\right)\, .
\end{equation}
The next task therefore is the discussion of the integrand 
as function of $n$ and $\alpha$ in order to determine the conditions for the convergence of the integral. We can check this for both factors
independently for their behaviour at the lower and upper limits of the integral. The algebraic factor $1/u^{2(1-\alpha)}$
is convergent for $u \rightarrow 0$ if $\alpha > 1/2$.
The exponential factor is convergent for $s \rightarrow \infty$
if $H_0 > 0$ and $n > 1$. If we swap the behaviours and require the algebraic term to ensure converge at large values of $u$, we
need $\alpha < 1/2$. Requiring the exponential term to ensure
convergence for small $u$, we find the inverted criterion defined by
$ n < 1$ and $ H_0 < 0$. In the case of these convergence conditions, we can write the result in the expression 
\begin{equation}
\label{inteval}
\frac{1}{K} = \frac{1}{|n-1|}\left(\frac{n-1}{H_0}G_0^2\right)^\frac{2\alpha - 1}{n-1} \Gamma\left(\frac{2\alpha - 1}{n-1}\right)\, , 
\end{equation}
with the Gamma function $\Gamma(x)$. 
This result is curious as it yields perfectly normalizable distributions on the ``It\^o-side", $\alpha < 1/2$ (with $ n < 1$ and $ H_0 < 0$), and on the
``anti-It\^o side", $ \alpha > 1/2$ (with $H_0 > 0$ and $n > 1$), of the discretization parameter interval $0 \leq \alpha \leq 1$, while the usually most physically relevant Stratonovich value is left out! Figure 1
displays the result for $P_{\infty}(u)$ for the It\^o and
anti-It\^o cases for a number of chosen parameters.
Let us recall the physical meaning of the parameters $\alpha$, $H_0$, and $n$.
The coefficient $\alpha$ represents the discretisation parameter, and therefore the type of stochastic integral used. It defines the position of the point at
which we calculate any integrated stochastic function in the small
intervals of the adopted Riemann sum ($0\le \alpha\le1$). The coefficient $H_0$ controls the intensity of the drift, and its sign discriminates between dissipative forces ($H_0<0$), and active or non-dissipative forces ($H_0>0$). The coefficient $n$ represents the type of nonlinearity, which shows a singularity at the origin when $n<0$, a root-type behavior when $0<n<1$, and a singularity at infinity when $n>1$. Finally, the density can be normalised on the ``It\^o-side" when the forces are dissipative and the non-linearity is of the root or divergent type at the origin; it can be normalised on the ``anti-It\^o side" when the force is active with singularities at infinity.

\begin{figure}[t]
\centering\includegraphics[width=2.5in]{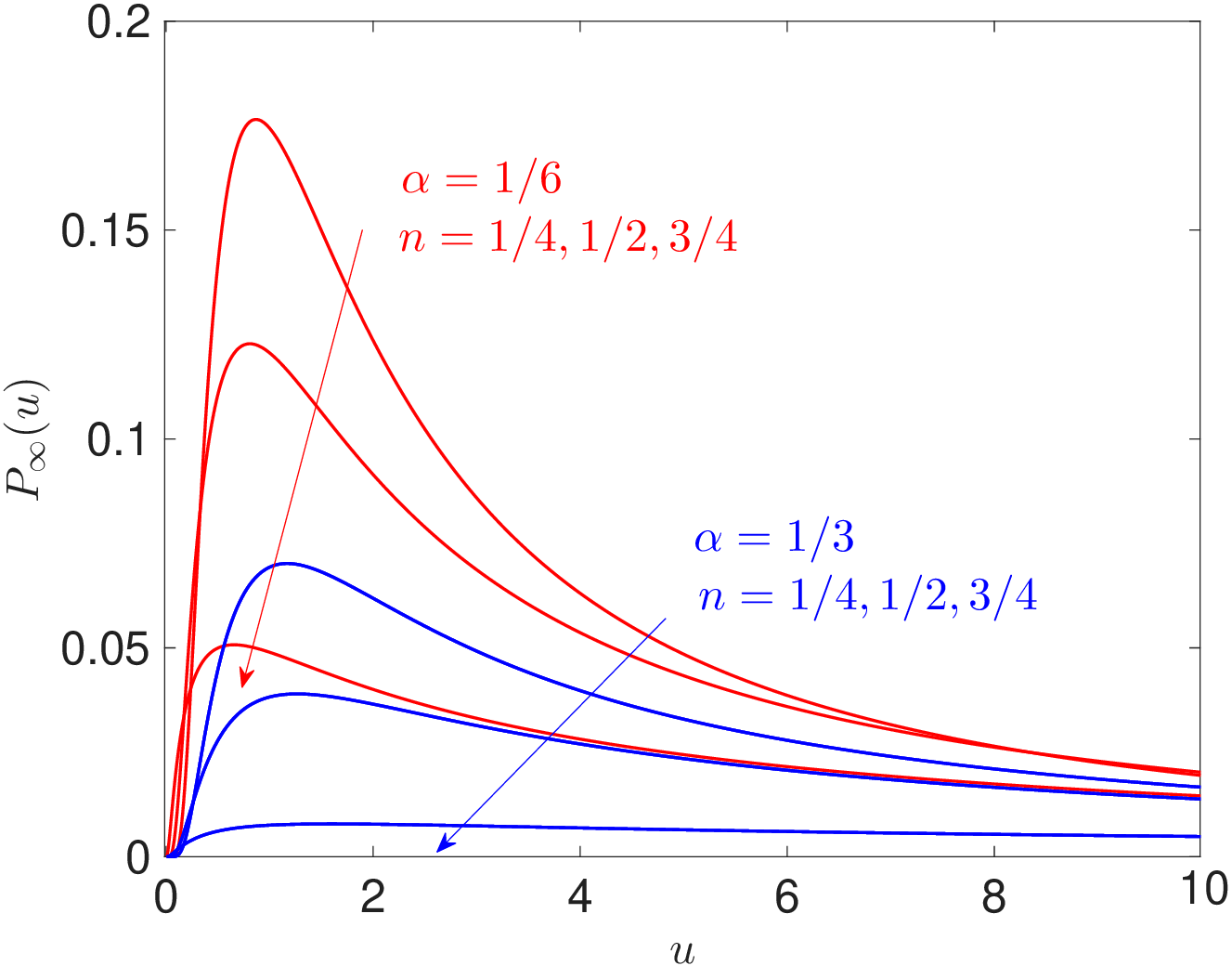}
\centering\includegraphics[width=2.4in]{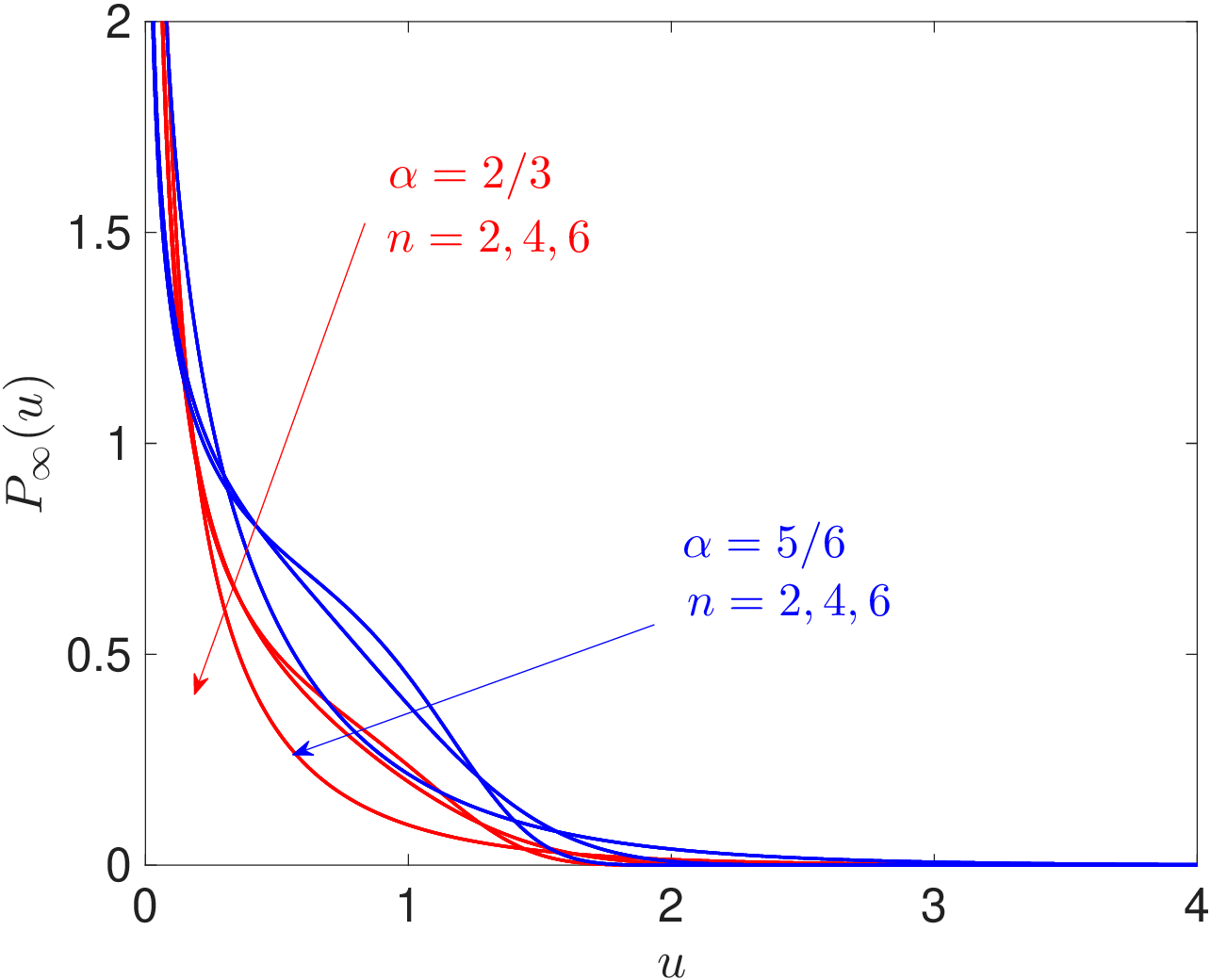}
\caption{Left: $P_{\infty}(u)$ on the ``It\^o-side" $\alpha < 1/2$ (with $ n < 1$ and $ H_0 < 0$);
$H_0 = -3/2$ and $G_0 = 1$; Right: $P_{\infty}(u)$ on the ``anti-It\^o side", $\alpha > 1/2$ (with $H_0 > 0$ and $n > 1$), with $H_0 =3/2$ and $G_0 = 1$.
}
\label{fig_sim}
\end{figure}

To conclude this discussion, we observe that eq. (\ref{langGGBM}) can exhibit blow-up phenomena (the solutions tend to
infinity as the variable $s$ approaches a finite value) only in particular cases that we have not considered in our development. Indeed, by using the Feller test for explosions \cite{karatzas91,mck,cherny}, it is possible to prove that eq. (\ref{langGGBM}) exhibits
blow-up phenomena if and only if $H_0 < 0$ and
$n > 1$, regardless of the stochastic interpretation adopted (for details see also Ref.\cite{giordano23}). 

\section{Infinite Ergodicity and the Stratonovich Case}

We have seen that the case $\alpha = 1/2$ does not lead to a
normalizable stationary PDF for a nonlinear drift term in the 
previous section. There is, however, an approach with allows to
make sense of this situation: this is the infinite ergodicity approach, which we briefly introduce here, following \cite{leibovich19,aghion20,giordano23}.
In contrast to the pedagogical approach chosen in these cited papers which takes recourse
to classical problems in statistical mechanics, we here develop the argument straightaway from the Langevin and Fokker-Planck equations we wrote down for
the slightly modified geometric Brownian motion, i.e. 
\begin{equation}
    \frac{du}{ds} = -H_0 u^n + G_0 u \xi(s)\, , 
\end{equation}
considering the case $H_0 < 0$ and $n < 1$ 
(for which there can be no blowup phenomena, as mentioned above). 
Hence the asymptotic PDF, following our earlier result, reads as
\begin{equation}
    P_{\infty}(u) \sim \frac{1}{u} \exp{\left(- \frac{H_0}{G_0^2}\frac{u^{n-1}}{n - 1}\right)}\,. 
\end{equation}
The second piece of information we have is the exact solution of the
Fokker-Planck equation with $\alpha=1/2$ in the absence of the drift term, $H_0 = 0$,
\begin{equation}
    P(u,s) = \frac{\exp\left(-\frac{\log(u/\overline{u}_0)^2}{4G_0^2s}\right)}{2uG_0
    \sqrt{\pi s}}
\end{equation}
for the initial condition $P(u,0) = \delta(u - \overline{u}_0)$. We suppose that for
large scales $s \rightarrow \infty$, {\it in the presence of the drift $H_0$},
the resulting distribution is given by
\begin{equation}
P(u,s) \underset{s \to \infty}\sim
\frac{\exp\left(-\frac{\log(u/\overline{u}_0)^2}{4G_0^2s}\right)}{2uG_0
    \sqrt{\pi s}}\exp{\left(- \frac{H_0}{G_0^2}\frac{u^{n-1}}{n - 1}\right)}\, ,
\end{equation}
which for large scales becomes
\begin{equation}
P(u,s) \underset{s \to \infty}\sim
\frac{1}{2uG_0 \sqrt{\pi s}}
    \exp{\left(- \frac{H_0}{G_0^2}\frac{u^{n-1}}{n - 1}\right)}\, .
\end{equation}
It can be verified by direct computation that this putative solution
indeed fulfills the Fokker-Planck equation
\begin{equation}
    \frac{\partial}{\partial s}P(u,s) =
    H_0 \frac{\partial}{\partial u}(u^n P(u,s))+
    G_0^2\frac{\partial}{\partial u}
    \left(u\frac{\partial}{\partial u}(uP(u,s))\right)\,,
\end{equation}
in the limit of large scales. This means specifically that all terms of order $s^{-1/2}$ in the equation cancel exactly, leaving only one remaining term of order $s^{-3/2}$, which becomes negligible in the limit as it decays
more readily.

We now postulate the expression 
\begin{equation} \label{inv}
    {\cal I} \equiv \lim_{s \rightarrow \infty} 2G_0\sqrt{\pi s} P(u,s) =
    \frac{1}{u} \exp{\left(- \frac{H_0}{G_0^2}\frac{u^{n-1}}{n - 1}\right)}
\end{equation}
as an {\it invariant density} in the sense that the expression for $P(u,s)$
found above can be used to compute expectation values of physical observables via
\begin{equation}
\label{average1}
     \langle{ {\cal O} (u) }\rangle (s) = \int_0^{\infty} du\, {\cal O}(u)P(u,s)\, .
\end{equation}
From the introduction of the invariant density $\mathcal{I}$, the asymptotic behavior of the average value $\langle{ {\cal O} (u) }\rangle (s)$ can be deduced by the following explicit  expression
\begin{equation}
\lim_{s\to\infty}2G_0\sqrt{\pi s}\left\langle \mathcal{O}(u) \right\rangle (s)=\int_{0}^{+\infty}du\frac{\mathcal{O}(u)}{u}\exp\left(-\frac{H_0}{G_0^2}\frac{u^{n-1}}{n-1} \right).
\label{average2}
\end{equation}

\begin{figure}[ht]
\centering\includegraphics[width=4in]{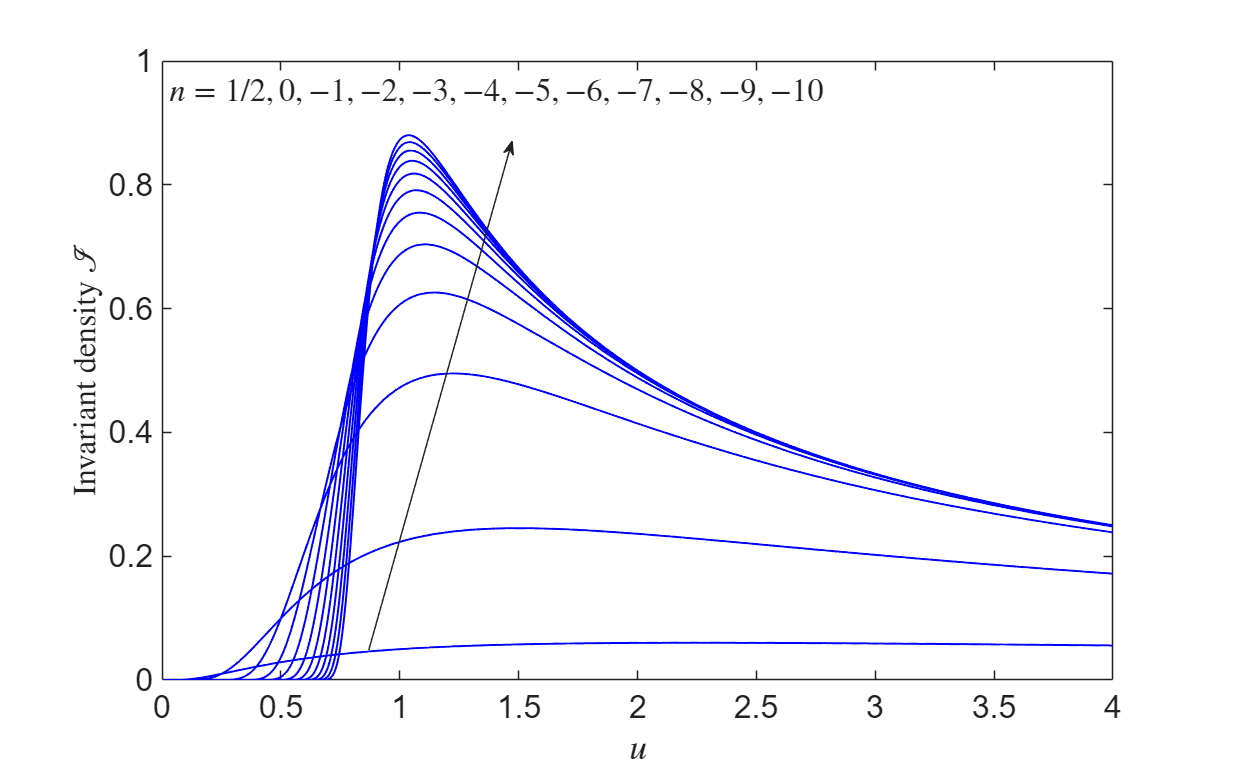}
\caption{The invariant density $\mathcal{I}$ from eq.  (\ref{inv}) with different values of $n$. We adopted $H_0=-3/2$ and $G_0=1$.
}
\label{fig_2}
\end{figure}

If, e.g., we take as observable the power law ${\cal O}(u) = u^{n-1} $, the result is given by
\begin{equation}
\label{limlim}
    \lim_{s \rightarrow \infty} 2G_0\sqrt{\pi s} \langle u^{n-1} \rangle(s) 
    = \frac{G_0^2}{|H_0|}\, ,
\end{equation}
i.e., the ratio of the diffusion and drift coefficients. We chose the observable $ u^{n-1} $ because it represents the force when we write the Langevin equation in the form $\frac{1}{u}\frac{du}{ds} = -H_0 u^{n-1} + G_0  \xi(s)$. However, this result can be easily generalized to consider an arbitrary observable ${\cal O}(u) = u^{s}$. In this case, the integral in eq.  (\ref{average2}) is analogous to that defined in eq.  (\ref{partinte}), provided that we substitute $2(1-\alpha)$ with $1-s$. Hence, the previous study of the integral allows us to state that $\lim_{s\to\infty}2G_0\sqrt{\pi s}\left\langle \mathcal{O}(u) \right\rangle (s)$ is finite if $s<0$. Its value can be obtained through eq.  (\ref{inteval}), with $2(1-\alpha)$ substituted with $1-s$. When we let $s=n-1$ (which is negative), we recover eq.  (\ref{limlim}).

Finally, we can move on to consider, in the same spirit, the presence of a nonlinear diffusion term in the Langevin equation which will then be given by
\begin{equation}
    \frac{du}{ds} = - H_0u^n + G_0u^m\xi(s)\, .
    \label{langgen}
\end{equation}
We can follow the same reasoning as before and end up with the
asymptotic PDF given by
\begin{equation}
    P_{\infty}(u) = \frac{K}{u^{2m(1-\alpha)}}
    \exp\left(-\frac{H_0}{G_0^2}\frac{u^{n-2m+1}}{n-2m +1}\right)\, ,
\end{equation}
and the analogous expression for the finiteness of $1/K$ as before, now
with the additional parameter $m$. So we need to see how the conditions
we obtained before for $m=1$ are now modified for $m\neq 1$. 

The reasoning is essentially identical to what we presented before, since
one has to regularize the divergent behaviours at the integration limits.
The cases i) and ii) from before are replaced by
\begin{equation} \label{cond1}
    2m(1 - \alpha) < 1\,,\,\,\, H_0 > 0\,,\,\,\, n - 2m + 1 > 0,
\end{equation}
\begin{equation} \label{cond2}
    2m(1 - \alpha) > 1\,,\,\,\, H_0 < 0\,,\,\,\, n - 2m + 1 < 0\, ,
\end{equation}
respectively. The conditions 
(\ref{cond1}) and (\ref{cond2}) can be combined to
\begin{equation}
    \frac{G_0^2}{H_0}\Delta(n,m) > 0
\end{equation}
and
\begin{equation}
    \frac{2m\alpha - 2m + 1}{\Delta(n,m)} > 0
\end{equation}
with $\Delta(n,m) \equiv n-2m + 1$.

The normalization constant $K$ can be calculated in these
parameter regions and the result is expressed as
\begin{equation} \label{pdf_nm}
    P_{\infty}(u) = 
\frac{
|\Delta(n,m)| 
\exp\left(
-\frac{H_0}{G_0^2}
\frac{u^{\Delta(n,m)}}{\Delta(n,m)}\right)
}
{\left(\Delta(n,m)\frac{G_0^2}{H_0}\right)^{\frac{2m\alpha - 2m + 1}{\Delta(n,m)}}
\Gamma\left(\frac{2m\alpha - 2m + 1}{\Delta(n,m)}\right)u^{2m(1-\alpha)}
}\, .
\end{equation}
\\
Figure 3 displays the asymptotic PDF for  negative and positive values of the amplitude
$H_0$ (which means for both regions defined in eqs.  (\ref{cond1}) and (\ref{cond2})).

\begin{figure}[ht]
\centering\includegraphics[width=2.4in]{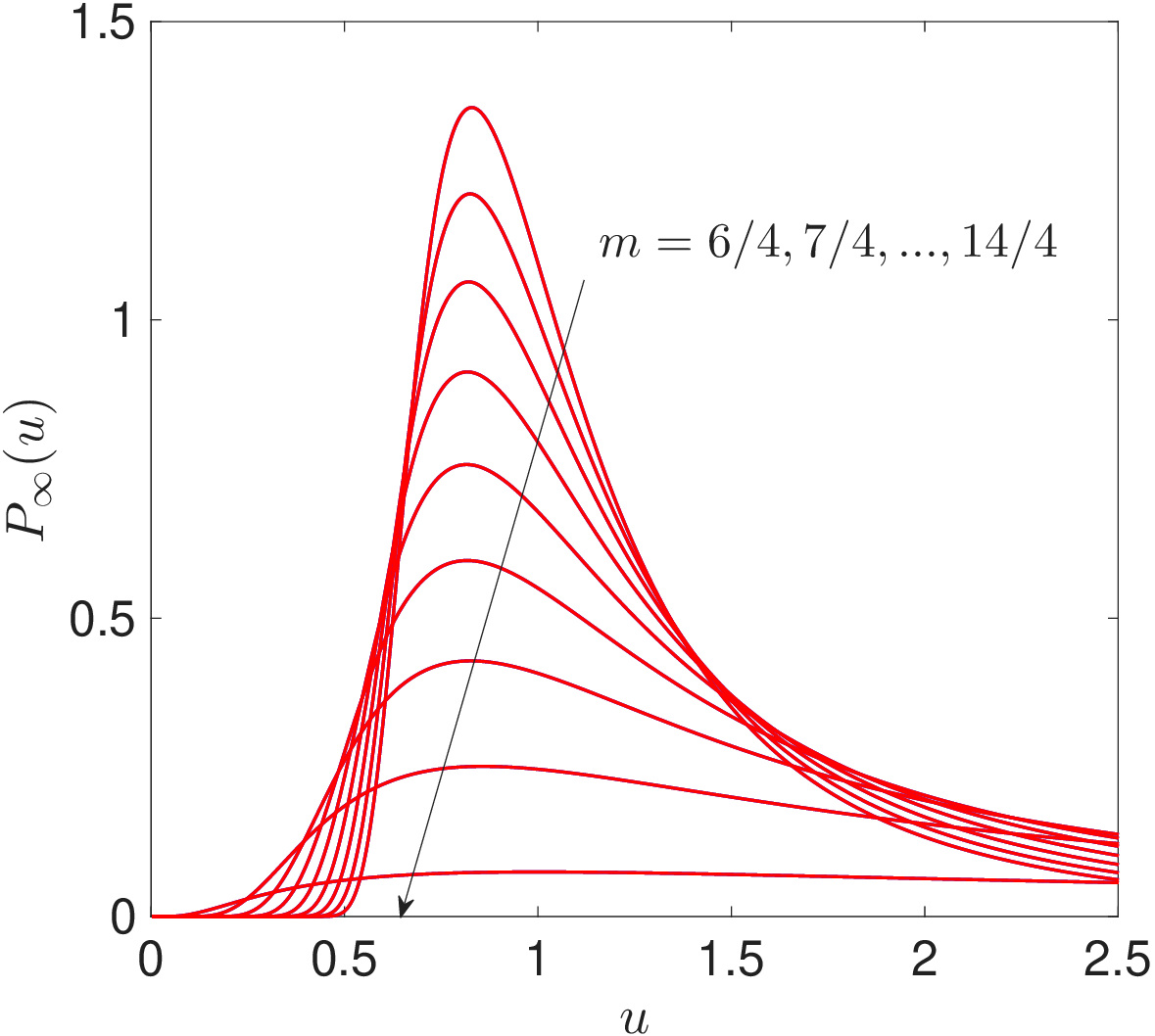}
\centering\includegraphics[width=2.4in]{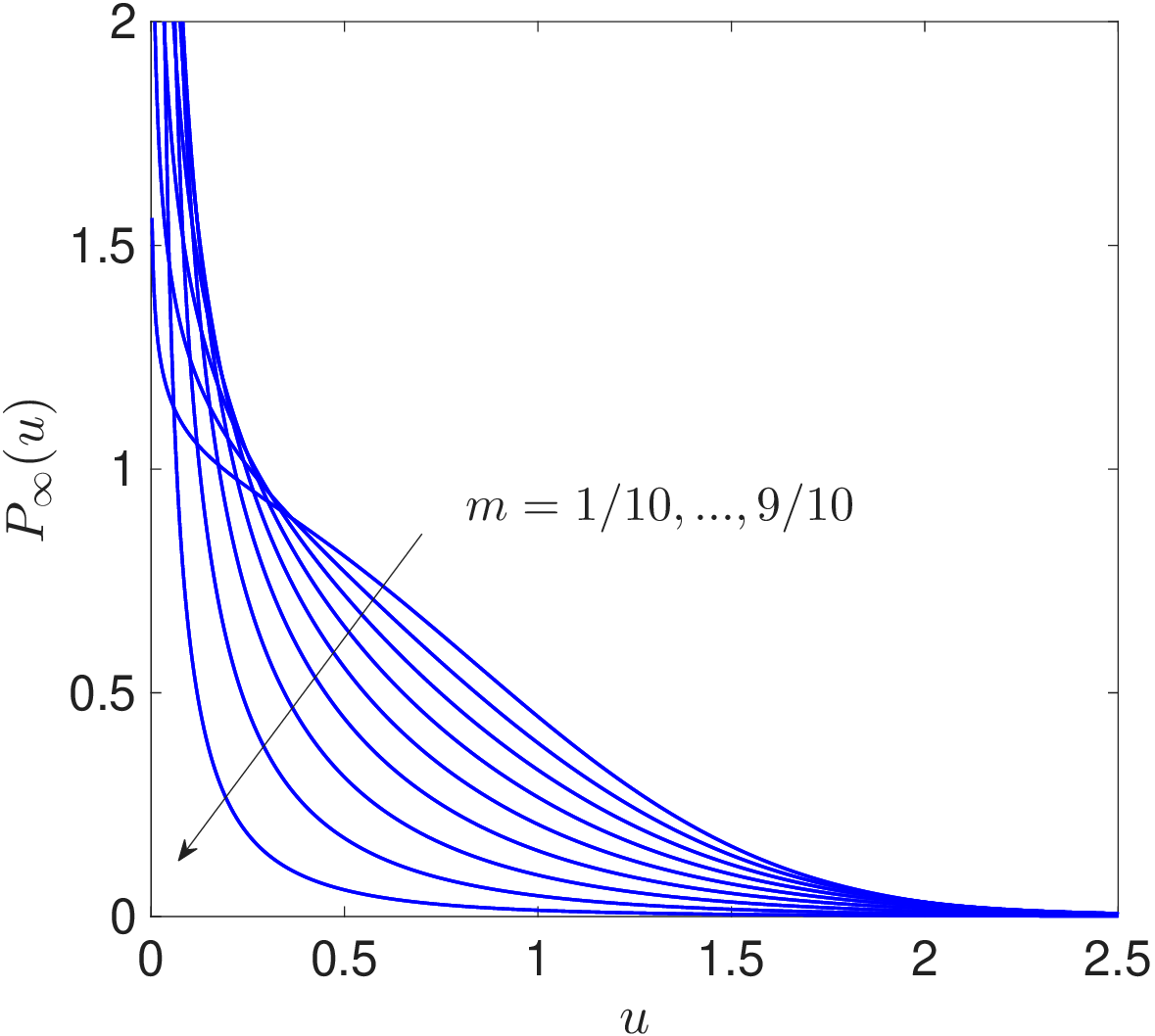}
\caption{Left: $P_{\infty}(u)$ from eq. (\ref{pdf_nm}) with eq. (\ref{cond2}) satisfied. We adopted $H_0 = - 3/2$,
$G_0=1$, $\alpha=1/2$, and $n=3/2$. Right: $P_{\infty}(u)$ from eq. (\ref{pdf_nm}) with eq. (\ref{cond1}) satisfied. We adopted $H_0 =  3/2$,
$G_0=1$, $\alpha=1/2$, and $n=3/2$. The values of $m$ are specified in both panels. }
\label{fig3}
\end{figure}

It is important to mention that also for eq. (\ref{langgen}) blow-up or explosion phenomena can be observed. The Feller test can be applied to this equation and it is possible to prove that we have explosions in finite time
if and only if $H_0<0$, $n > 2m-1$, and $n > 1$ (or, equivalently $n > max\left\lbrace2m - 1, 1\right\rbrace$),  regardless of the stochastic interpretation adopted. Therefore we cannot have blow-up phenomena in our two regions defined in eqs.  (\ref{cond1}) and (\ref{cond2}).

It is interesting to note that if we consider the particular case where $n=1/3$ and $m=2/3$, we recover the Obukhov-Richardson theory of turbulence \cite{sawford06,skvortsov10}. In the Richardson theory, it is easy to see that the variance of the process evolves as $t^3$ \cite{richardson26,barkai12,barkai14}. Note that in this case our parameter $\Delta(n,m) \equiv n-2m + 1$ is zero and therefore the asymptotic density cannot be normalized within this theory. Nevertheless, the following application of infinite ergodicity cannot be applied to this case because $\Delta(n,m)=0$  is not allowed, as we will see below. This theory remains applicable if we allow for small differences in the coefficients $n$ and $m$.

In the final step we can treat, as before, the infinite ergodicity construction based on solution of the 
time-dependent Fokker-Planck equation without drift, i.e., the solution for the nonlinear diffusion. 
For $H_0=0$ and $m \neq 1$ it has been given in the literature already before \cite{leibovich19} in the form
\begin{eqnarray} \label{P-H0}
        P(u,s)  = 
    \frac{\overline{u}_0^{\frac{1}{2}(1 - 2m\alpha)}u^{\frac{1}{2}(1 - 4m + 2m\alpha)}}
    {2G_0^2(1-m)s} 
    \exp{\left(-\frac{\overline{u}_0^{2(1-m)}+u^{2(1-m)}}{4G_0^2(1-m)^2s}\right)}
    I_{\nu}\left(\frac{\overline{u}_0^{1-m} u^{1-m}}{2G_0^2(1-m)^2s}\right) ,
\end{eqnarray}
for the initial condition $P(u,0) = \delta(u - \overline{u}_0)$, and where
$I_{\nu}$ is the Bessel function of the first kind of order $\nu = (1-2m\alpha)/(2(m-1))$\,. 
The solution is correct for $0\leq m< 1$ and $2m\alpha-2m+1>0$.
For $\alpha = 1/2$, one has $I_{-1/2}(z) = 
(2/(\pi z))^{1/2} \cosh(z)$, and the expression can then be written in
the symmetrized form
\begin{equation} \label{Pnm}
    P(u,s) = \frac{1}{2G_0\sqrt{\pi s} u^m}\sum_{k=-1,1}
    \exp{\left(-\frac{\left(u^{1-m} + k\overline{u}_0^{1-m}\right)^2}{4G_0^2(1-m)^2s}\right)}\, .
\end{equation}
This expression describes the superposition of an incident density for 
$ k = -1 $ generated at $u = \overline{u}_0$ and a reflected density for $k = +1$.
For $m = 0$, incident and reflected densities become Gaussians, in line with
the noise being additive in this case. For $\alpha \neq 1/2$, the symmetry between incident and reflected solutions is broken. 

Finally, we can repeat the steps that led us to our previous result (\ref{inv}). From the asymptotic behaviour of the Bessel function
for $ z \rightarrow 0 $
\begin{equation} 
        I_{\nu}(z) \sim \frac{1}{\Gamma(\nu+1)}\left(\frac{z}{2}\right)^{\nu} ,
\end{equation}
and from eq. (\ref{P-H0}) we can obtain the asymptotic behaviour of the full
solution
\begin{eqnarray}
\label{solgeneasy}
P(u,s)\underset{s \to \infty}\sim\frac{[2(1-m)]^{(1-2\alpha)\frac{m}{1-m}}}{(G_0^2s)^{\frac{2m\alpha-2m+1}{2(1-m)}}\Gamma\left[ \frac{2m\alpha-2m+1}{2(1-m)}\right] u^{2m(1-\alpha)}}.
\end{eqnarray}
In the Stratonovich case $\alpha = 1/2$ the result simplifies to
\begin{equation}
    P(u,s) \underset{s \to \infty}\sim \frac{1}{G_0\sqrt{\pi s}u^m} .
\end{equation}

For the equation without forcing term, 
$H_0=0$, there is no equilibrium and we know the asymptotic evolution when $0\leq m<1$ and $2m\alpha-2m+1>0$.
The aim of the infinite ergodicity is to give meaning to the solution of eq. (\ref{langgen}) even when it cannot be normalized. 
Hence, we consider the conditions $0\leq m<1$ and $2m\alpha-2m+1>0$, 
and we add the assumptions $H_0 < 0$ and $\Delta(n,m)=n-2m+1<0$, in such a way that $P_{\infty}(u)$ it is not normalizable. 

When this set of conditions is satisfied, we can merge Eqs.(\ref{pdf_nm})  and (\ref{solgeneasy}) in order to get the asymptotic behavior
\begin{eqnarray}
\label{solgeneasyfull}
P(u,s)\underset{s \to \infty}\sim\frac{[2(1-m)]^{(1-2\alpha)\frac{m}{1-m}}\exp\left(-\frac{H_0}{G_0^2}\frac{u^{\Delta(n,m)}}{\Delta(n,m)} \right)}{(G_0^2s)^{\frac{2m\alpha-2m+1}{2(1-m)}}\Gamma\left[ \frac{2m\alpha-2m+1}{2(1-m)}\right] u^{2m(1-\alpha)}},
\end{eqnarray}
and if $\alpha=1/2$, we  have
\begin{eqnarray}
\label{solstraasyfull}
P(u,s)\underset{s \to \infty}\sim\frac{\exp\left(-\frac{H_0}{G_0^2}\frac{u^{\Delta(n,m)}}{\Delta(n,m)} \right)}{G_0\sqrt{\pi s} u^m}.
\end{eqnarray}
These results represent the asymptotic behavior of $P(u,s)$ when $0\leq m<1$, $2m\alpha-2m+1>0$,  $H_0<0$, and $n-2m+1<0$. 
The formal proof can be obtained by direct substitution into the Fokker-Planck equation.

These asymptotic behaviors lead to the definition of the following invariant densities

\begin{eqnarray}
{\cal I}_{\alpha,n,m} =\lim_{s\to\infty}\frac{\Gamma\left[ \frac{2m\alpha-2m+1}{2(1-m)}\right](G_0^2s)^{\frac{2m\alpha-2m+1}{2(1-m)}}P(u,s)}{[2(1-m)]^{(1-2\alpha)\frac{m}{1-m}}}
=\frac{\exp\left(-\frac{H_0}{G_0^2}\frac{u^{\Delta(n,m)}}{\Delta(n,m)} \right)}{ u^{2m(1-\alpha)}}, 
\label{invm}
\end{eqnarray}
and  for $ \alpha = 1/2$ 
\begin{equation}
{\cal I}_{\frac{1}{2},n,m} =\lim_{t\to\infty}G_0\sqrt{\pi t}P(u,s)=\frac{\exp\left(-\frac{H_0}{G_0^2}\frac{u^{\Delta(n,m)}}{\Delta(n,m)} \right)}{ u^m}.
\label{inv12}
\end{equation}
Of course, these expressions can be used to determine the asymptotic behavior of the average value of physical observables, as already described in eqs.  (\ref{average1}) and (\ref{average2}). 

\section{The square-root processes}

\begin{figure}[ht]
\centering\includegraphics[width=4.5in]{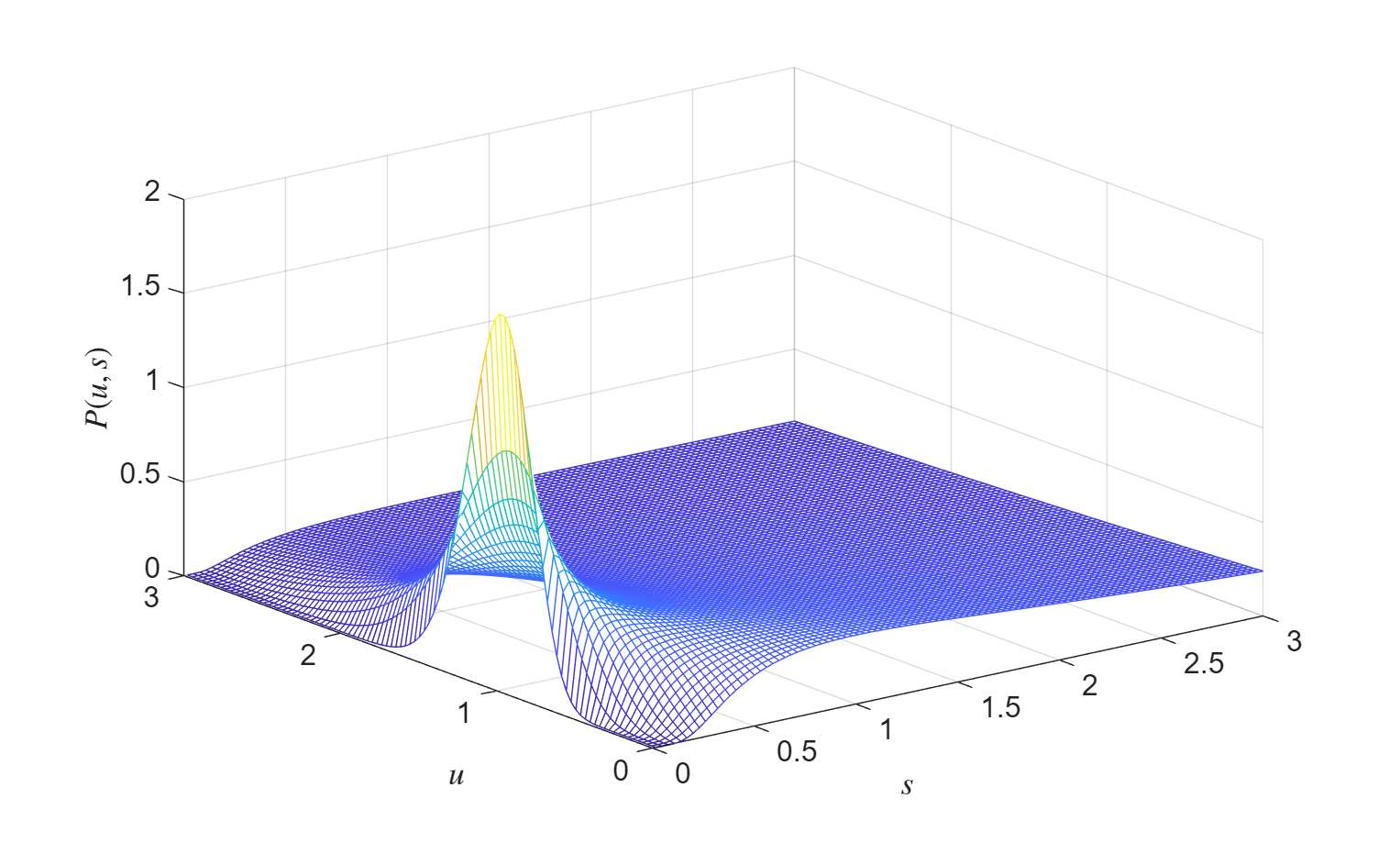}

\caption{Plot of the function $P(u,s)$ reported in eq.  (\ref{P-H0-tris}) with $\overline{u}_0=1$, $G_0=1$. We assumed that $0<u<3$ and $0.02<s<3$. The initial condition corresponds to  
$P(u,0) = \delta(u - \overline{u}_0)=\delta(u - 1)$. The solution corresponds to reflecting boundary conditions for $u=0$.}
\label{fig4}
\end{figure}

Although GBM has become the standard model for asset prices in mathematical finance and turbulence, its limitations have long been recognized, particularly when the linear multiplicative noise leads to unrealistically large fluctuations.
To overcome these difficulties, alternative diffusion processes were proposed in the 1970s and 1980s that replaced the linear scaling with sublinear forms \cite{karatzas91,revuz99}. 
Among the most important is the square-root diffusion (related to the Bessel process), in which the noise amplitude scales with the square root of the state variable. A landmark contribution came from Cox, Ingersoll, and Ross (1985), who introduced what is now known as the Cox-Ingersoll-Ross (CIR) process as a model of interest rates in finance \cite{cox85}.  Its importance rapidly extended beyond finance to areas such as population dynamics, queueing theory, neuroscience, and more recently, turbulence.
Another similar process is the so-called Heston process which is described by the coupling of a standard GBM with a square root or CIR process \cite{heston93}.  
Our process defined in eq. (\ref{langgen}), with $m=1/2$ assumes the square-root form
\begin{equation}
    \frac{du}{ds} = - H_0u^n + G_0\sqrt{u}\xi(s)\,,
    \label{langgenbis}
\end{equation}
similar to the processes just described, but with a nonlinear drift. If $H_0=0$, from eq. (\ref{P-H0}), we have the exact solution 
\begin{eqnarray} \label{P-H0-bis}
        P(u,s)  =\frac{1}
    {G_0^2s}  \left(\frac{\overline{u}_0}{u}\right)^{\frac{1-\alpha}{2}}
        \exp{\left(-\frac{\overline{u}_0+u}{G_0^2s}\right)}
    I_{\alpha-1}\left(2\frac{\sqrt{\,\overline{u}_0 u}}{G_0^2s}\right) ,
\end{eqnarray}
which for $\alpha=1$ reduces to
\begin{eqnarray} \label{P-H0-tris}
        P(u,s)  =\frac{1}
    {G_0^2s}  
        \exp{\left(-\frac{\overline{u}_0+u}{G_0^2s}\right)}
    I_{0}\left(2\frac{\sqrt{\,\overline{u}_0 u}}{G_0^2s}\right).
\end{eqnarray}
It is interesting to note that eq. (\ref{P-H0-tris}) is used in Ref.\cite{bensoussan16} to model the statistics of the two-dimensional velocity norm of a turbulent flow. It means that the  H\"anggi-Klimontovich (or anti-It\^o) interpretation is relevant to the study of the kinetic energy of turbulent motions. The plot of this function can be found in Fig. \ref{fig4}. We consider now the case with drift where $H_0<0$ and $n<0$. From eq. (\ref{solgeneasyfull}), we obtain the asymptotic behavior of the probability density as
\begin{eqnarray}
\label{solgeneasyfullbis}
P(u,s)\underset{s \to \infty}\sim\frac{\exp\left(-\frac{H_0}{G_0^2}\frac{u^n}{n} \right)}{(G_0^2s)^{\alpha}\Gamma\left( \alpha\right)u^{1-\alpha}},
\end{eqnarray}
which becomes for $\alpha=1$
\begin{eqnarray}
\label{solgeneasyfullbis1}
P(u,s)\underset{s \to \infty}\sim\frac{\exp\left(-\frac{H_0}{G_0^2}\frac{u^n}{n} \right)}{G_0^2s}.
\end{eqnarray}
These results can be used to define the invariant densities and to study the asymptotic evolution of average values of interest for square-root processes modified by nonlinear drift. This can be done for different drift exponents $n$ in order to possibly fit experimental data. 

In the context of turbulence, square-root diffusions have gained attention as models for quantities that must remain nonnegative and that exhibit strong fluctuations. A prime example is the instantaneous turbulent kinetic energy (TKE) or the local dissipation rate \cite{bensoussan16,bossy22,rolland18,pablo20}. Standard GBM would permit unphysical negative values in the presence of additive noise corrections, while purely multiplicative GBM often produces overly explosive dynamics. By contrast, the CIR-type process ensures positivity while moderating fluctuations through its sublinear noise scaling \cite{bensoussan16,bossy22,rolland18,pablo20}.
Phenomenological turbulence theories often require stochastic descriptions of cascade variables, intermittency corrections, or subgrid-scale dynamics in large-eddy simulations. Square-root diffusions provide a natural candidate because they encode intermittent bursts while preserving bounded statistical structure. Recent works in computational fluid dynamics have employed CIR-type models to represent the evolution of unresolved turbulent energy in Lagrangian stochastic frameworks \cite{bossy22}. These models can reproduce non-Gaussian statistics of velocity increments, heavy-tailed probability distributions, and scale-dependent intermittency \cite{yann18}. Moreover, the flexibility of the CIR drift term allows fitting to experimental or numerical data, capturing mean-reverting tendencies observed in turbulent flows.

Another line of application relates to anomalous transport and dispersion in turbulence \cite{rtis23,stani23}. Here, square-root diffusions serve as simplified proxies for energy or tracer densities, enabling analytical progress in problems where full Navier–Stokes dynamics are intractable. Their non-negativity, ergodic properties, and gamma-distributed stationary states resonate with the statistical features of dissipation fields observed in laboratory and numerical studies.

\section{Conclusions}

In this work, we have discussed algebraic generalizations of geometric Brownian motions (GGBMs). 
We motivated this study by phenomenological
statistical theories for turbulence, rather than the more commonly 
used stochastic processes in finance, for which questions of ergodicity
are somewhat less intuitive. Within a statistical physics approach, this
obviously is very different, and the question whether a stochastic process
has an invariant density or measure is of prime interest. Geometric Brownian
motion, as researchers have shown, can be used from fits to experimental data to be made consistent with the K62 scaling theory by Kolmogorov and
Obukhov. Since experimental suffer from unavoidable noise, the usually assumed functional dependence of the drift and diffusion terms on velocity
increments may be questioned -- at least with the purpose of testing the
robustness of the standard result. This in our view thus justifies to consider nonlinear generalizations of the geometric Brownian motion process
in order to understand their properties.

In the present work, we have looked at mildly nonlinear versions of a
generalized geometric Brownian motion, in which a linear drift and a linear
diffusion term is rendered algebraically nonlinear; one might think of 
just very moderate differences from linearity. Furthermore, we have considered a general value for the discretization of the stochastic integral, commonly described by a parameter $0 \leq \alpha \leq 1/2$. Our findings can be summarized as follows:

1) Already by considering an algebraic nonlinear drift term with power $n\neq 1$, a major difference arises between what we call the It\^o- and anit-It\^o regimes, with $0 < \alpha < 1/2$ and $1/2 < \alpha \leq 1$, respectively. Asymptotic distributions exist
in both regimes, with very different behaviours. At the Stratonovich value 
$\alpha = 1/2$, a conventional asymptotic distribution or invariant density does not exist.

2) In the Stratonovich case $\alpha = 1/2$, an invariant density can be
formulated by invoking the concept of infinite ergodicity. In this approach,
an asymptotic solution is constructed from the knowledge of the asymptotic 
density for $ n < 1 $ and the knowledge of the exact density of the diffusion process. Following this recipe, an invariant density can be
derived for large scale structures that can be used to calculate expectation values of physical quantities of interest.

This approach can also be made to work in the presence of a nonlinearity
$m \neq 1$ in the algebraic diffusion law. We have given the result for 
nonlinear drift $n \neq 1$ and $m\neq 1$ and arbitrary value $\alpha$,
within the limits for these parameters that make mathematical sense. 
As a special case, we have considered the nonlinear diffusion case with $m = 1/2$, the {\it square-root process}, which has recently found interest in
turbulent flows. 
Summing up, we hope that the recent developments of the concept of infinite ergodicity can find interest in more applications of physics and finance, notably in the field of turbulence.
Beyond what was discussed here, it might e.g. be worthwhile to study time averages instead of ensemble averages. Restart processes have recently shown the interest in such quantities
\cite{barkai23}.

\ack{We thank Tom Dupont and Rainer Grauer for discussions,
Fabrizio Cleri for a critical reading of the manuscript, and the referees for enlightening suggestions.}


\end{document}